\documentclass[runningheads]{llncs}
\usepackage[utf8]{inputenc}
\usepackage{url}
\usepackage{graphicx}
\usepackage{amssymb}

\title{Extended Reality for Smart Built Environments Design: Smart Lighting Design Testbed}
\titlerunning{Extended Reality for Smart Built Environments Design}
\author{Elham Mohammadrezaei \and
Denis Gra{\v{c}}anin\orcidID{0000-0001-6831-2818}}
\authorrunning{E. Mohammadrezaei, D. Gra{\v{c}}anin}
\institute{Virginia Tech, Blacksburg VA 24060, USA\\
\email{\{elliemh,gracanin\}@vt.edu}}

\begin{document}

\maketitle

\begin{abstract}
Smart Built Environment is an eco-system of `connected' and `smart' Internet of Things (IoT) devices that are embedded in a built environment.
Smart lighting is an important category of smart IoT devices that has recently attracted research interest, particularly for residential areas.
In this paper, we present an extended reality based smart lighting design testbed that can generate design prototypes based on the functionality of the physical environment.
The emphasis is on designing a smart lighting system in a controlled residential environment, with some evaluation of well-being and comfort.

\keywords{Smart lighting  \and Design \and Internet of Things \and Extended Reality.}
\end{abstract}

\section{Introduction}

Smart built environment (SBE) or smart home provides an eco-system of `connected' and ‘smart’ Internet of Things (IoT) devices situated in the built environment~\cite{Handosa-2020-a}.
The main goal of SBE is to provide and promote user comfort, quality of living, convenience and security to satisfy residents' needs.
To help achieve this goal, we build on our previous results~\cite{Dasgupta-2019-b,Gracanin-2015-c,Handosa-2020-a} to provide an SBE design testbed where designers can use the designs prescribed by the testbed as well as to explore original designs.

The Internet has fundamentally altered our way of life, allowing people to engage virtually in a variety of settings ranging from work to social relationships.
By facilitating communications with and among smart objects, IoT has the potential to add a new layer to this process, resulting in the idea of ``anytime, anywhere, anymedia, everything'' communications~\cite{Atzori-2010-a}.

New sensor, mobile, and control technologies have a lot of potential for connecting people with their surroundings.
Individuals, groups, and the broader community can benefit from smart built environments (e.g., a smart house) that are enhanced with technology, such as increased awareness of information in the user's surroundings, integrated control over factors in one's surrounding and home environments, and increased ability to support sustainable living for both individuals and groups.

IoT~\cite{Atzori-2010-a} is a concept that defines the ubiquitous presence of things or devices that utilize a unique addressing system to interact with one another and collaborate with their neighbors to achieve common goals.
These physical items have a social existence that IoT might sustain.
IoT has a wide range of applications, from automation and manufacturing to assisted living and e-health.
The ability to change how systems behave and how users interact with them is provided by designing and deploying IoT into built settings.
Design, simulation, planning, monitoring, optimization, and visualization technologies could all help with sustainability~\cite{Gracanin-2015-c}.

IoT has penetrated the daily operations of numerous industries as a result of recent improvements in communication and mobile computer technology; applications include, but are not limited to, smart agriculture, smart grids, smart buildings, and e-health.

Smart buildings are an important part of smart cities, and they've been the subject of a lot of research in recent years.
Despite the fact that a vast variety of infrastructures, platforms, and systems have been developed, implemented, and deployed, several barriers such as upfront technology investment and continuing system maintenance costs have stymied their widespread adoption.

The equipment is often purchased by the owners of these solutions, who are also responsible for its implementation and upkeep.
The inherent hazards have become one of the most significant impediments to wide-scale adoption of IoT technology.
Furthermore, a great number of past systems have only been installed and tested in a single building or a testbed that only covers a few levels, therefore their availability and scalability in large-scale real buildings are unclear~\cite{8710297}.

Smart lighting is a novel concept that has emerged over a decade now and is being used and tested in commercial and industrial built environments.
Currently, smart lighting research is predominantly dedicated to energy saving in non-residential environments; the residential environments have not been explored~\cite{soheilian2021smart}.
The focus is mostly on designing and developing a smart lighting system in a controlled environment, with a limited evaluation of well-being and comfort.

In this paper, we focus on designing and developing an Extended Reality (XR) based SBE smart light design testbed that is capable of generating design solutions and alternatives based on the functionality of the space.

\section{Related Work}



Several smart home systems based on IoT applications have recently been developed with the goal of making human life more convenient and ecologically friendly.
Real-world situations, on the other hand, present a number of difficulties.
A smart home is built to be energy efficient and provide basic functionality such as lighting and switch settings.

Furthermore, a low-cost network can be created by combining embedded controllers, such as Arduino with Ethernet, ZigBee technology, and an Android device that acts as a home environment controller.
The system's drawbacks don't apply to all security solutions, and such a solution isn't new in a smart home~\cite{lee2019effects,zhihua2016design}.

There are methods for evaluating IoT-based built environments that employs a large-scale virtual environment (VE) in which a building model is aligned with the physical space~\cite{Gracanin-2015-c}.
To model user interaction with constructed spaces, this approach takes advantage of affordances and embodied cognition in a vast physical setting.

The model is based on an enclosed space with real-time tracking and spatial audio capabilities.
Several users can move around the physical space at the same time and view the simulation results from various perspectives.
The corresponding view in the VE is determined in real-time based on a user's physical location and orientation.
The user can also monitor and cooperate with other users to change the simulation.


A smart home system employs integrated sensors, actuators, wireless networks, and graphical user interfaces to provide positive, adaptable, secure, and cost-effective results.
By incorporating sensors for lighting, temperature, pressure, humidity, motion, fire alarms, and dust/air, among other things, a sensor network may transform an existing home into a smart home~\cite{bhatt2016cost}.
However, this platform faces numerous challenges, one of which is the issue of security and privacy.

Similarly,~\cite{7906845} provides a simple platform based on open-source code, in which the authors present a solution for remote monitoring in a smart home by using ESP8266 micro-controller and MQTT protocol~\cite{MQTT-2015}.
This minimizes the IoT system's security risk while also increasing its cost.




\subsection{Extended Reality for Smart Built Environments Design}

The architectural, engineering, and construction (AEC) industry has increasingly recognized XR technology for its ability to provide multisensory three-dimensional (3D) environments that immerse the user in a virtual world, specifically to meet the high demand for visual forms of communication during the work related to designing, engineering, construction, and management of the built environment over the past decades~\cite{kim2013virtual}.

The first application of XR technology for the built environment can be traced back to the 1990s, when this simulation technology was first brought to the attention of architects, piqued the interest of other disciplines involved in architectural engineering, and led to further exploration of XR technology's possibilities.

Recent advances in computer graphics and virtual reality equipment have resulted in a slew of useful XR applications in pilot testing and industry, including client walkthroughs, review, and building sequence visualization.
AEC professionals have increasingly pushed and implemented XR technology to many other disciplines, leveraging on its sophisticated functions such as visualization, in addition to these traditional methods.

It's worth noting that XR might refer to a simulation of specific features of the real world, a symbolic world, or an imaginative world, among other things.

XR technology and associated applications have been progressively investigated and deployed in the AEC industry because to its huge potential.
Nonetheless, such technology has yet to be embraced as a standard tool in the workplace.
According to market reports, user experience (i.e., cumbersome gear and technological faults), content offering, and cost are among the challenges preventing XR from becoming popular~\cite{perkins2020augmented}.

As a summary for discussion, challenges and opportunities of XR applications for the built environment, potential research needs, such as 1) user-centered adaptive design, 2) human cognition-driven virtual reality information system, 3) construction training system incorporating human factors, 4) occupant-centered facility management, and 5) industry adoption, are proposed to shed light on future research directions~\cite{zhang2020virtual}.

\subsection{Light}

In home automation, smart lighting is one of the main components.
Integrating it into the Smart House platform is a fundamental part of using the benefits that new technologies give us in everyday life.
The use of centralized lighting control contributes to more efficient use of electricity and adds an extra dose of comfort in the home.

Xu et al. proposed a smart construction framework that solution providers can use as a starting point when designing their own solution architecture. Following that, they present a smart lighting system for smart buildings based on the framework.
In their system, the standard emergency lights can be augmented to a smart router by simply replacing the old product with a low-cost wireless module~\cite{8710297}.


It's worth noting that in recent years, an increasing number of researchers have worked on improving control algorithms.
Methods such as statistics, data modeling, and machine learning have been utilized to save energy.
A neural network controller, for example, was created and tested.
It may regulate the brightness of bulbs in a classroom based on the ambient light and the number of students~\cite{chen2013artificial}.

Besides saving energy, some studies also considered to improve user experience~\cite{byun2013intelligent,may2013smart,parise2013design}.
For consumers' convenience and a better experience, some researchers supported integrated lighting control with occupancy sensors, photocells, and a central control module~\cite{middleton2013integrated}.
A few researchers developed mobile applications for better user experience in terms of operability and mobility~\cite{castillo2018evaluation,choi2016dynamic,moon2016implementation,suresh2016automatic}.
Evidence suggests that a well-designed mobile application can enhance user experience while simultaneously ensuring compliance with lighting rules and lowering energy use~\cite{castillo2018evaluation}.

Every lighting designer, architect and project manager should have suitable tools to plan their project in the best way possible.
Table~\ref{fig:dataset1} shows a comparative study of the features of different packages that are suitable for use in lighting design.

\begin{table}[t]
\centering

\caption{\label{fig:dataset1}
Comparative analysis of different lighting simulation packages.
 }
\scriptsize
\begin{tabular}{|p{1in}|c|c|c|c|c|c|c|c|} \hline 
\textbf{\small Feature} & AGI 32 & CalcuLux &DIALux & Radiance & MicroLux & LightCalc & Visual 3D & Unity \\ \hline
Interior and exterior study & \checkmark & \checkmark & \checkmark & & & & \checkmark & \checkmark \\ \hline
Analysis and visualization of lighting design & & & & \checkmark & & & & \\ \hline
Designing in 2D and 3D organize & & & & & \checkmark & & & \checkmark \\ \hline
Irregular geometries & \checkmark & & & & & & & \checkmark \\ \hline
Analysis of the appropriate distances between objects & & & & & & \checkmark & \checkmark & \checkmark \\ \hline
Analysis and comparison of different lighting scenarios & & \checkmark & & \checkmark & & & \checkmark & \checkmark \\ \hline
Calculation tools & & \checkmark & & & & & \checkmark & \\ \hline
QuickTime Virtual reality (QTVR) & & & & & & & & \checkmark \\ \hline
Virtual Reality Markup Language (VRML) & & & & & & & & \checkmark \\ \hline
Walkthrough animation & & & & & & & & \checkmark \\ \hline
\end{tabular}
\end{table}

\section{Problem Description}

There are many intelligent lighting control strategies, but they are all based on basic lighting control methods: on-off control and dimming control.
Constant illuminance control refers to using luminance management in conjunction with intelligent sunshade to keep the illuminance within a consistent range.
When an infrared sensor detects that someone has entered the room, the luminaire will automatically turn on, while an illumination sensor detects the room's illumination and adjusts the luminaire's brightness, keeping the room's illumination close to the default.
At the same time, manage the shading blind's rotation angle automatically to minimize direct glare and excessive indoor temperatures.
The lighting qualities of lamps produce varied levels of illumination in different areas of a room.


Timing control refers to the use of a timer to switch on and off a luminaire. 
Timing control can be used in the bedroom, for example.
Turn on luminaire automatically at a specific time in the morning to assist people in waking up as quickly as possible; in the evening, turn off luminaire automatically at a given time to encourage people to go to bed as quickly as possible. 

A combination of timer and illumination sensor can manage lighting in places like balconies and green plants.
The landscape lighting in the region is automatically turned on when the illumination falls below a specified threshold during a certain time period; otherwise, it is turned off.
As a result, it's both attractive and functional, as well as energy-efficient and environmentally beneficial.

Intelligent lighting now not only offers light to people but also promotes a healthy lifestyle.
Scene control refers to the ability to turn on different luminaires based on different conditions in order to provide the desired lighting atmosphere. 
People must do a wide range of activities in a relatively short space due to the peculiarities of dwelling.
The lighting requirements for various behaviors aren't always the same.
People can watch TV, read, chat, and sip tea in the living room, for example.
To create a comfortable, energy-efficient lighting atmosphere, scene control can be used to initiate multiple scene illumination schemes.


By studying and forecasting people's behavior, linkage control refers to automatically turning luminaires on and off in a relevant zone. 
When individuals open the closet, for example, the luminaire in it will automatically turn on for picking clothes, and when they close it, the luminaire will turn off.
When individuals sit in front of a dresser, the mirror light will turn on and stay within a predefined illumination range; in the meantime, the luminaire will turn off automatically when they depart.
When people get up in the middle of the night, the nightlight in their bedroom will automatically switch on.
The luminaires in the respective region will automatically switch on and maintain a reasonably low-light standard when people enter the living room, bathroom, and other rooms, and will automatically turn off when people leave.

To produce an intelligent, pleasant lighting environment, linkage control enhances and strengthens existing management measures.
These control strategies are, in reality, accomplished on the basis of on-off and dimming control, and they are not mutually independent.
A good intelligent lighting system should develop a technologically reasonable, cheap, practical, healthy, and comfortable intelligent lighting management scheme by considering all control techniques in the context of actual applications.

Intelligent lighting offers the following benefits as compared to traditional lighting.
In intelligent lighting, lighting in various environments is precisely and intelligently controlled by various means, and minimum amounts of energy are used to ensure required illumination standards, effectively avoiding the phenomena of ``everlasting lamps'' and overly strong illumination while also drastically reducing energy consumption. According to statistics, intelligent lighting may save more than 30\% of energy.

A decent working or living environment requires a proper lighting environment.
Intelligent lighting makes use of advanced electronic and information technology to automatically control the on-off and brightness of lamps in order to increase uniformity of illumination and reduce the stroboscopic effect, resulting in more comfortable and healthy lighting environments for people's work and lives, as well as reducing dizziness and eyestrain.

There is a strong need for a comprehensive framework for the smart lighting design as a part of the  SBE design process.
As a result, smart capabilities of lighting design are underutilized in terms of improving occupants' overall spatial experience and influencing activity patterns.
We looked at existing systems from a variety of domains to help us establish a smart lighting design system for smart home design.
Our proposed system stands in the intersection of full manual and full automatic control where the user will be provided with embedded design ideas that are automatically suggested based on the geometry and its functionality and in the next step, the user will be able to manually change the settings and modify them based on their needs, activities and the weather condition and probably many other factors.



\begin{figure}[t]
\centering
\includegraphics[width=\linewidth]{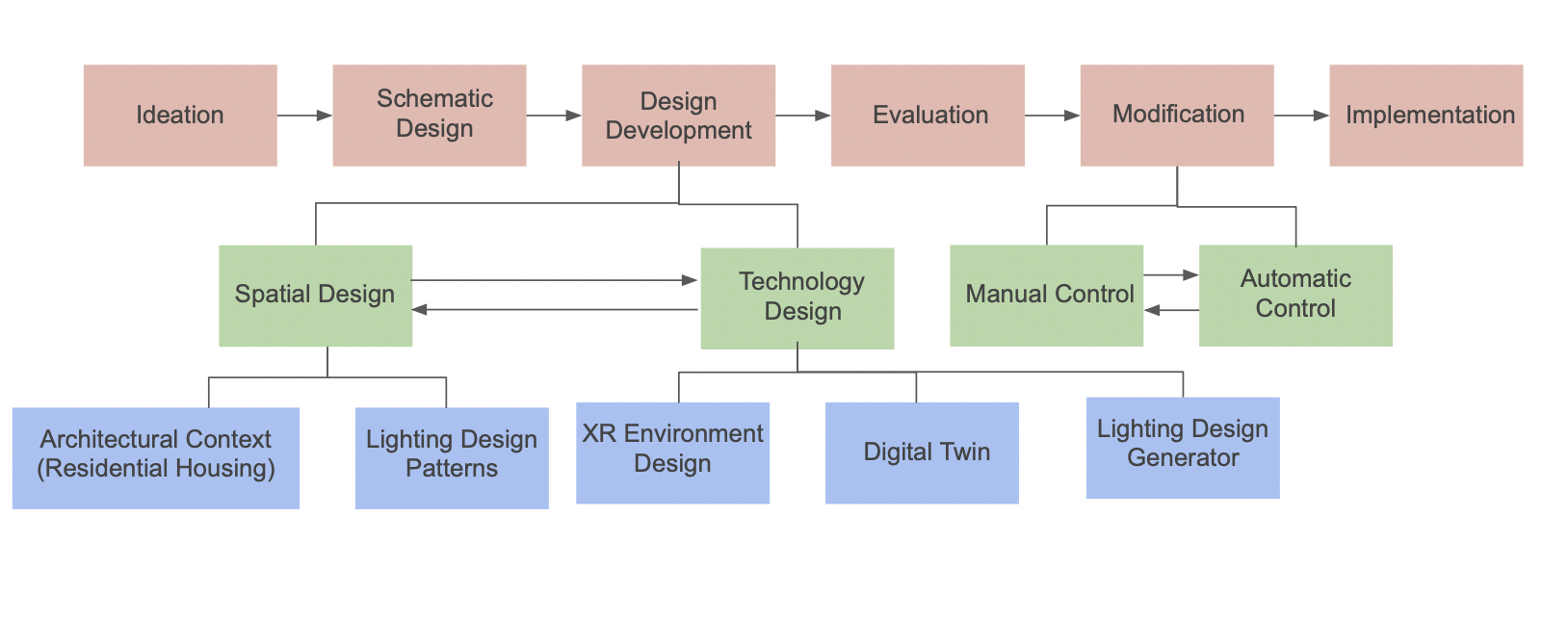}
\caption{\label{fig:prototype1}
Design process diagram.
}
\end{figure}

\section{Proposed Approach}
We studied SBE research to understand the architectural concerns for smart home design.
We also reviewed smart lighting design for SBE and identified the components and the current research focus to identify the existing smart lighting systems and the role of light in a successful architectural design and how the smartness in lighting design can help increasing the quality of users' experience of their living spaces.

Additionally, we reviewed XR technology and its application to architecture and construction industry and its ability to create multi-sensory and three dimensional environments which provides an opportunity for users to experience and interact with a virtual version of the future building and modify it based on their needs and fix the flaws in pre-construction phase.

To investigate the current state, challenges, and best practices of the smart lighting design process, we studied past researches in lighting design and existing intelligent lighting system and how these systems add smartness to the system.
We also studies about current software package which are being used for simulating the lighting systems and had a quick review of the features of each software and what are the pros and cons of each option.

Smart capabilities are often underutilized in terms of improving occupants' overall spatial experience and influencing activity patterns.
We explored design methods from a variety of domains to help us designing a smart lighting design generator system.

Our SBE smart lighting design testbed (Figure~\ref{fig:prototype1}) includes control, communication and interconnection capabilities and enables designers to have full control over light characteristics and how occupant control those lights (interaction modalities).
The proposed testbed targets a sweet spot between having a full manual and full automatic control over the lighting design and modifications which means at the very first step, given the input, it automatically generates and proposes the prototypes and later in the development process, it allows for manual changes and modifications by users.

\begin{figure}[t]
\centering
\includegraphics[width=\linewidth]{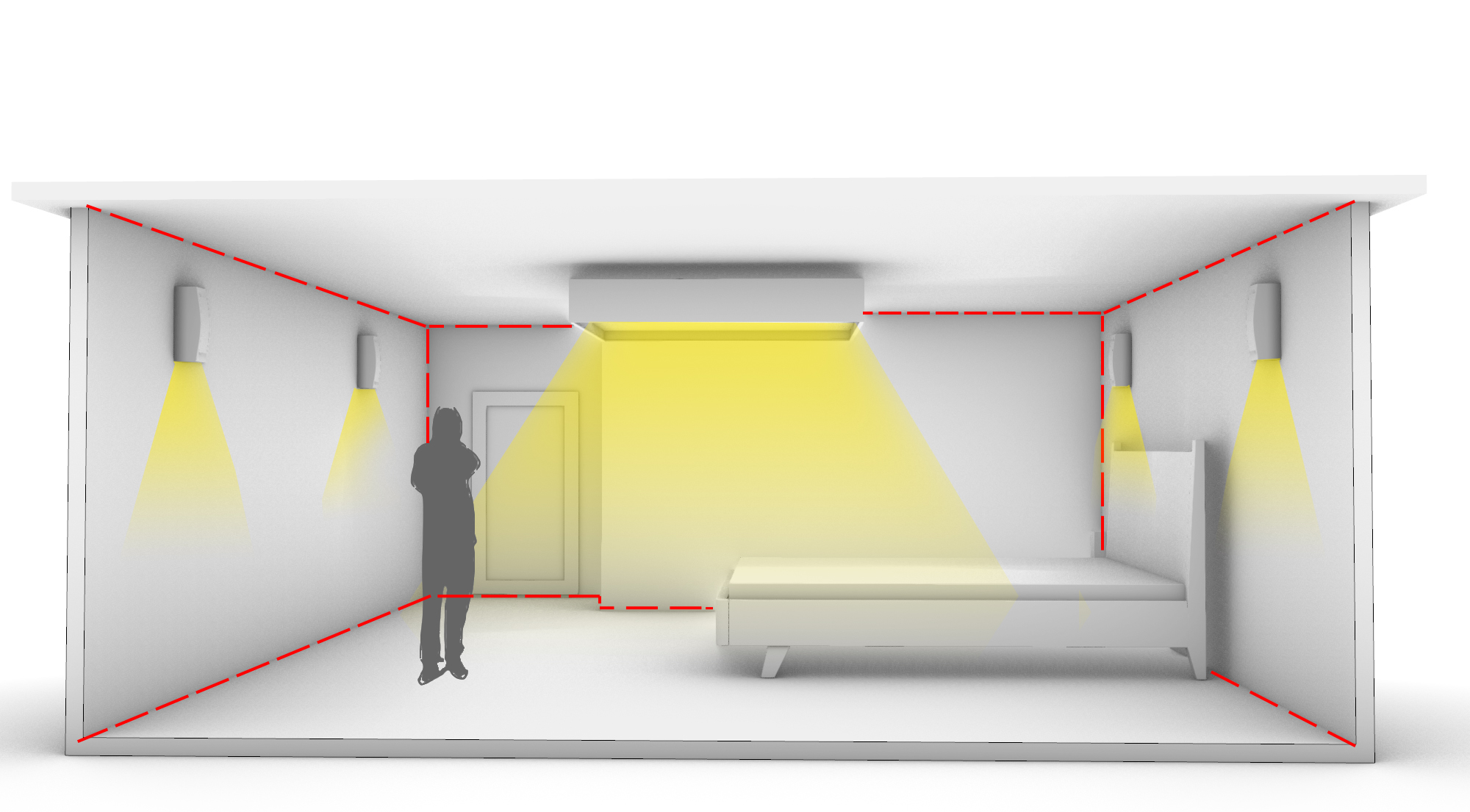}
\caption{\label{fig:prototype2}
Ideation diagram.
}
\end{figure}

Using cross-platform tools, we provide a visual representation of lighting design prototypes, as well as management and monitoring capabilities, manual control and modification in parallel with having automatic control on generating prototypes based on the geometry and also functionality of the input context.

The testbed is capable of understanding what the functionality of the space is and propose design prototypes based on the guidelines and use design patterns to provide an initial design without user intervention.

The testbed includes a collection of lighting design patterns and outlines of different room types which are developed according to the U.S. Interior and Exterior Lighting Systems and Controls Guideline~\cite{LightingGuideline} and are embedded in the system.
For example, a design pattern for a bedroom in residential housing provides a solution that includes a ceiling mounted luminaire which provides ceiling surface brightness as well as two table light sources that provide task lighting.

\section{Case Study}

\begin{figure}[t]
\centering
\includegraphics[width=\linewidth]{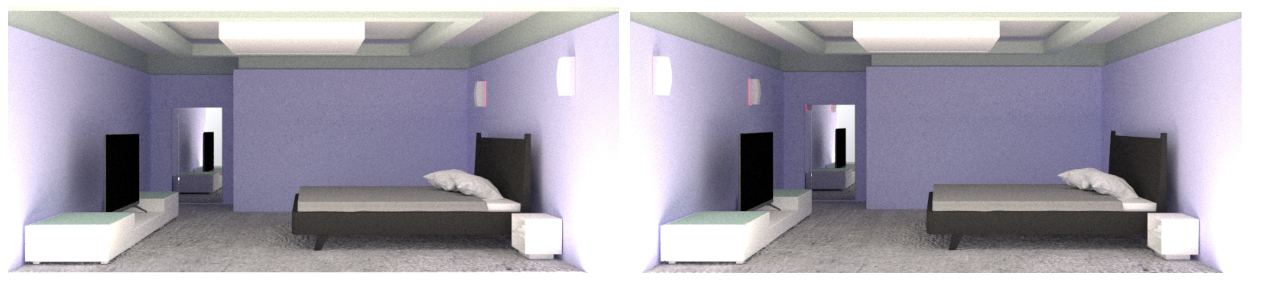}
\caption{\label{fig:prototypeA}
Comparative analysis of different lighting scenarios:
\textbf{Left:} ceiling lamp and
\textbf{Right:} wall mounted lamps on two sides of bed and/or TV.
 }
\end{figure}
\begin{figure}[t]
\centering
\includegraphics[width=\linewidth]{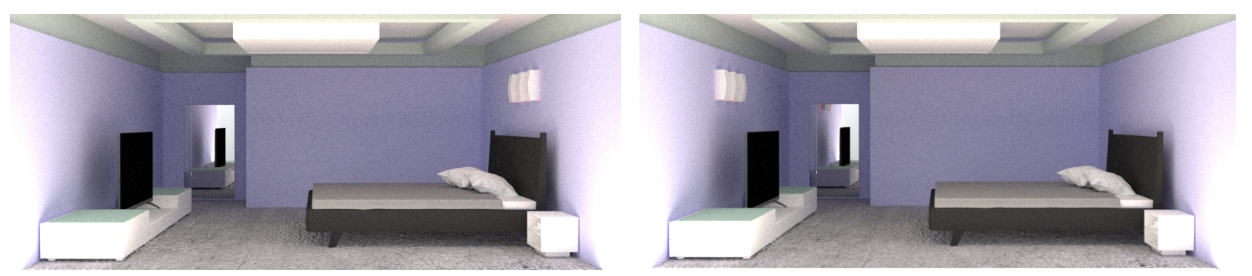}
\caption{\label{fig:prototypeB}
Comparative analysis of different lighting scenarios:
\textbf{Left:} ceiling lamp and
\textbf{Right:} wall mounted lamp on top of bed and/or TV.
 }
\end{figure}
\begin{figure}[t]
\centering
\includegraphics[width=0.925\linewidth]{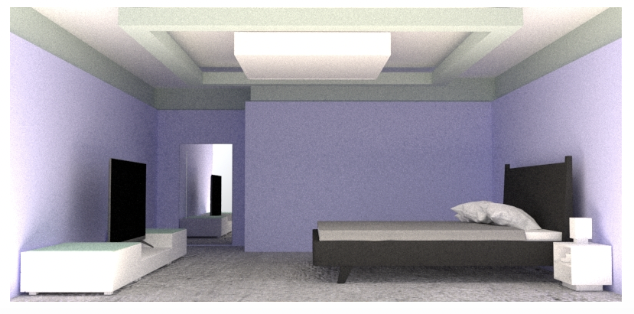}
\caption{\label{fig:prototypeC}
Comparative analysis of different lighting scenarios: ceiling lamp.
}
\end{figure}

For our case study we focus on a simple bedroom outline from a residential unit since it's one of the most important sections in a house and has direct influence on the residents' well-being and preservation of their comfort.
We design the system to generate lighting design prototypes and let the users get involved with the environment and have interaction with the system and evaluate it and give us feedback.
A sample bedroom outline has been designed and used as the context of this study and the proposed system has been implemented in Unity.
Figure~\ref{fig:prototype2} is a diagram that shows how the system first scans the room and figures out the outline and identifies the geometry and objects inside the room and in the next step it proposes situating the lighting fixtures inside that geometry and based on the function we have already assigned to the room.
Five different lighting design scenarios (Figures~\ref{fig:prototypeA},~\ref{fig:prototypeB}, and~\ref{fig:prototypeC}) have been proposed by system.

\begin{figure}[t]
\centering
\includegraphics[width=0.925\linewidth]{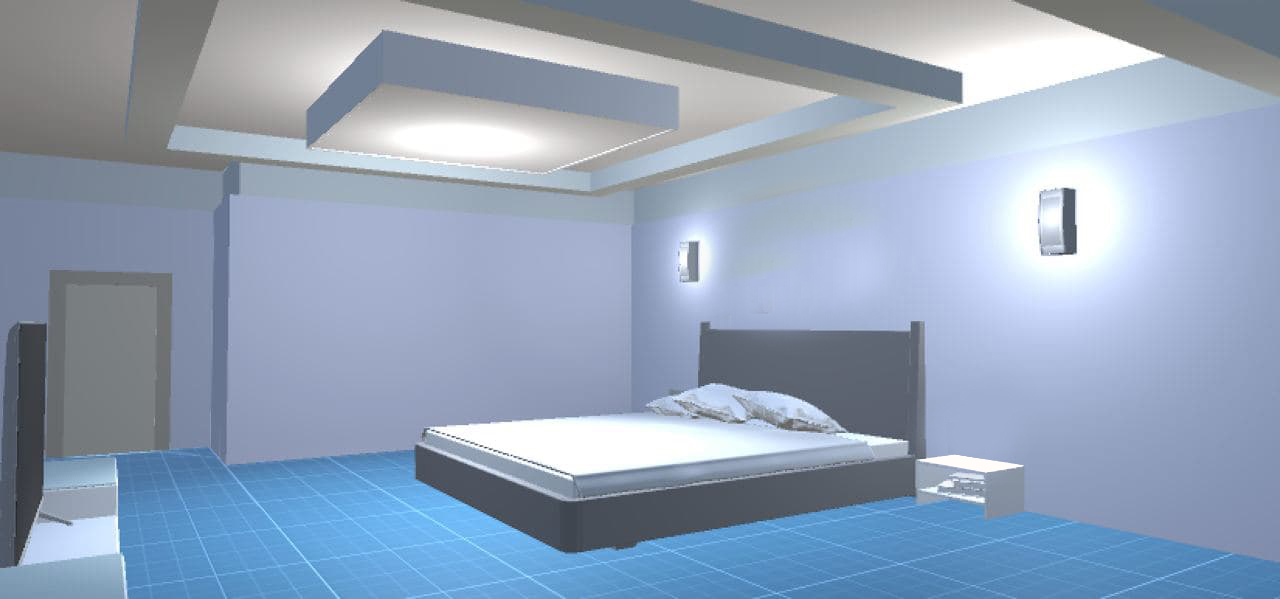}
\caption{\label{fig:prototypeD}
Real time lighting design rendering of Figure~\ref{fig:prototypeA} left.
}
\end{figure}

One of the corresponding rendering alternative is shown in Figure~\ref{fig:prototypeD}.


\section{Discussion}

Our SBE smart lighting design testbed includes control, communication and interconnection capabilities and enables designers to have full control over light characteristics and how occupant control those lights (interaction modalities). 

The initial design solution could be modified by a designer using the SBE testbed.
The designer can explore how a smart lighting system in an SBE responds to occupants’ behavioural habits and how a smart lighting system can have impacts on the occupants, how it can change the habits and also what are the effects on the occupants from a psychological perspective.

The described SBE smart lighting design testbed represents just one of the aspects of the SBE design.
Therefore, our future work will include expanding the testbed to include the support for other SBE functions and services.

Also, this system can be expanded to be implemented for other architectural contexts and environment rather than residential buildings and we also can develop the system and provide the outdoor and area lighting simulator too which can also generate design ideas for those spaces and so will be applicable to a broader range of buildings and functionalities.

\bibliographystyle{plain}
\bibliography{hcii_2022_light}

\begin{thebibliography}{10}

\bibitem{LightingGuideline}
{---}.
\newblock Unified facilities criteria ({UFC}): Interior and exterior lighting
  systems and controls.
\newblock Change 4 UFC 3-530-01, U.S. Department of Defense, 1~November 2019.

\bibitem{perkins2020augmented}
{---}.
\newblock Augmented and virtual reality survey report: Industry insights into
  the future of immersive technology.
\newblock Volume 4, Perkins Coie, March 2020.

\bibitem{MQTT-2015}
Martin Arlitt, Manish Marwah, Gowtham Bellala, Amip Shah, Jeff Healey, and Ben
  Vandiver.
\newblock {MQTT} version 3.1.1 plus errata 01.
\newblock Standard, OASIS, 10~December 2015.

\bibitem{Atzori-2010-a}
Luigi Atzori, Antonio Iera, and Giacomo Morabito.
\newblock The {I}nternet of {T}hings: A survey.
\newblock {\em Computer Networks}, 54(15):2787--2805, 2010.

\bibitem{bhatt2016cost}
Ashutosh Bhatt and Jignesh Patoliya.
\newblock Cost effective digitization of home appliances for home automation
  with low-power {W}i{F}i devices.
\newblock In {\em Proceedings of the 2nd International Conference on Advances
  in Electrical, Electronics, Information, Communication and Bio-Informatics
  (AEEICB)}, pages 643--648. IEEE, 2016.

\bibitem{byun2013intelligent}
Jinsung Byun, Insung Hong, Byoungjoo Lee, and Sehyun Park.
\newblock Intelligent household led lighting system considering energy
  efficiency and user satisfaction.
\newblock {\em IEEE Transactions on Consumer Electronics}, 59(1):70--76, 2013.

\bibitem{castillo2018evaluation}
Ana Castillo-Martinez, Jose-Amelio Medina-Merodio, Jose-Maria
  Gutierrez-Martinez, Juan Aguado-Delgado, Carmen de~Pablos-Heredero, and
  Salvador Ot{\'o}n.
\newblock Evaluation and improvement of lighting efficiency in working spaces.
\newblock {\em Sustainability}, 10(4):1110, 2018.

\bibitem{chen2013artificial}
Yifei Chen and Qian Sun.
\newblock Artificial intelligent control for indoor lighting basing on person
  number in classroom.
\newblock In {\em Proceedings of the 9th Asian Control Conference (ASCC)},
  pages 1--4. IEEE, 2013.

\bibitem{choi2016dynamic}
Kyungah Choi and Hyeon-Jeong Suk.
\newblock Dynamic lighting system for the learning environment: performance of
  elementary students.
\newblock {\em Optics express}, 24(10):A907--A916, 2016.

\bibitem{Dasgupta-2019-b}
Archi Dasgupta, Mohamed Handosa, Mark Manuel, and Denis Gra{\v{c}}anin.
\newblock A user-centric design framework for smart built environments: A mixed
  reality perspective.
\newblock In Norbert Streitz and Shin'ichi Konomi, editors, {\em Distributed,
  Ambient and Pervasive Interactions (International Conference on
  Human-Computer Interaction HCII 2019)}, volume 11587 of {\em Lecture Notes in
  Computer Science}, pages 124--143, Cham, 26--31~July 2019. Springer
  International Publishing.

\bibitem{Gracanin-2015-c}
Denis Gra{\v{c}}anin, Kre{\v{s}}imir Matkovi{\'{c}}, and Joseph Wheeler.
\newblock An approach to modeling {I}nternet of {T}hings based smart built
  environments.
\newblock In L.~Yilmaz, W.~K~V. Chan, I.~Moon, T.~M.~K. Roeder, C.~Macal, and
  M.~Rosetti, editors, {\em Proceedings of the 2015 Winter Simulation
  Conference (WSC)}, pages 3208--3209, 6--9~December 2015.

\bibitem{Handosa-2020-a}
Mohamed Handosa, Archi Dasgupta, Mark Manuel, and Denis Gra{\v{c}}anin.
\newblock Rethinking user interaction with smart environments---a comparative
  study of interaction modalities.
\newblock In Norbert Ztreitz and Shin'chi Konomi, editors, {\em Distributed,
  Ambient and Pervasive Interactions. HCII 2020}, volume 12203 of {\em Lecture
  Notes in Computer Science}, pages 39--57, Cham, 19--24~July 2020. Springer.

\bibitem{kim2013virtual}
Mi~Kim, Xiangyu Wang, Peter Love, Heng Li, and Shih-Chung Kang.
\newblock Virtual reality for the built environment: a critical review of
  recent advances.
\newblock {\em Journal of Information Technology in Construction},
  18(2013):279--305, 2013.

\bibitem{7906845}
Ravi~Kishore Kodali and SreeRamya Soratkal.
\newblock {MQTT} based home automation system using esp8266.
\newblock In {\em Proceedings of the 2016 IEEE Region 10 Humanitarian
  Technology Conference (R10-HTC)}, pages 1--5, 2016.

\bibitem{lee2019effects}
Hoonyong Lee, Changbum~R Ahn, Nakjung Choi, Toseung Kim, and Hyunsoo Lee.
\newblock The effects of housing environments on the performance of
  activity-recognition systems using {W}i-{F}i channel state information: An
  exploratory study.
\newblock {\em Sensors}, 19(5):983, 2019.

\bibitem{may2013smart}
Zazilah~Binti May and Yasir Ashiq Ali~B Mohd~Yaseen.
\newblock Smart energy saving classroom system using programmable logic
  controller.
\newblock In {\em Advanced Materials Research}, volume 660 of {\em Advanced
  Materials Research}, pages 158--162. Trans Tech Publications Ltd, 4 2013.

\bibitem{middleton2013integrated}
Stuart Middleton-White, Gregory Smith, Robert Martin, Thomas~J Hartnagel,
  Theodore~E Weber, Mike Crane, Terry Arbouw, Dawn~R Kack, and David~J Rector.
\newblock Integrated lighting system and method, May~7 2013.
\newblock US Patent 8,436,542.

\bibitem{moon2016implementation}
Seung-Mi Moon, Sook-Youn Kwon, and Jae-Hyun Lim.
\newblock Implementation of smartphone-based color temperature and wavelength
  control led lighting system.
\newblock {\em Cluster Computing}, 19(2):949--966, 2016.

\bibitem{parise2013design}
Giuseppe Parise, Luigi Martirano, and Giorgio Cecchini.
\newblock Design and energetic analysis of an advanced control upgrading
  existing lighting systems.
\newblock {\em IEEE Transactions on industry applications}, 50(2):1338--1347,
  2013.

\bibitem{soheilian2021smart}
Moe Soheilian, G{\'e}za Fischl, and Myriam Aries.
\newblock Smart lighting application for energy saving and user well-being in
  the residential environment.
\newblock {\em Sustainability}, 13(11):6198, 2021.

\bibitem{suresh2016automatic}
S~Suresh, HNS Anusha, T~Rajath, P~Soundarya, and SV~Prathyusha Vudatha.
\newblock Automatic lighting and control system for classroom.
\newblock In {\em Proceedings of the 2016 International Conference on ICT in
  Business Industry \& Government (ICTBIG)}, pages 1--6. IEEE, 2016.

\bibitem{8710297}
Weitao Xu, Jin Zhang, Jun~Young Kim, Walter Huang, Salil~S. Kanhere, Sanjay~K.
  Jha, and Wen Hu.
\newblock The design, implementation, and deployment of a smart lighting system
  for smart buildings.
\newblock {\em IEEE Internet of Things Journal}, 6(4):7266--7281, 2019.

\bibitem{zhang2020virtual}
Yuxuan Zhang, Hexu Liu, Shih-Chung Kang, and Mohamed Al-Hussein.
\newblock Virtual reality applications for the built environment: Research
  trends and opportunities.
\newblock {\em Automation in Construction}, 118:103311, 2020.

\bibitem{zhihua2016design}
Su~Zhihua.
\newblock Design of smart home system based on zigbee.
\newblock In {\em Proceedings of the 2016 International Conference on Robots \&
  Intelligent System (ICRIS)}, pages 167--170. IEEE, 2016.

\end{thebibliography}

\end{document}